\documentclass[aps,prb, amsfonts, amssymb, amsmath, superscriptaddress, notitlepage,twocolumn,10pt,nobalancelastpage]{revtex4-2}
\usepackage{graphicx} 
\usepackage{hyperref}
\usepackage{bm}
\hypersetup{ 
    colorlinks=true, 
    linktoc=all,     
    linkcolor=blue,  
    citecolor =blue
} 
\usepackage{amsmath}

\newcommand{\bk}{\mathbf{k}}
\newcommand{\bq}{\mathbf{q}}
\newcommand{\br}{\mathbf{r}}

\newcommand{\nn}{\nonumber}

\begin{document}

\title{Stability of the N\'eel quantum critical point in the presence of Dirac fermions}

\author{Huanzhi Hu}
\affiliation{London Centre for Nanotechnology, University College London, Gordon St., London, WC1H 0AH, United Kingdom}
\author{Jennifer Lin}
\affiliation{London Centre for Nanotechnology, University College London, Gordon St., London, WC1H 0AH, United Kingdom}
\author{Mikolaj D. Uryszek}
\affiliation{London Centre for Nanotechnology, University College London, Gordon St., London, WC1H 0AH, United Kingdom}
\author{Frank Kr\"uger}
\affiliation{London Centre for Nanotechnology, University College London, Gordon St., London, WC1H 0AH, United Kingdom}
\affiliation{ISIS Facility, Rutherford Appleton Laboratory, Chilton, Didcot, Oxfordshire OX11 0QX, United Kingdom}

\begin{abstract}
We investigate the stability of the N\'eel quantum critical point of  two-dimensional quantum antiferromagnets, described by a non-linear $\sigma$ model (NL$\sigma$M),
in the presence of a Kondo coupling to $N_f$ flavours of two-component Dirac fermion fields.  The long-wavelength order parameter fluctuations are subject to Landau damping by 
electronic particle-hole fluctuations. Using momentum-shell RG, we demonstrate that the Landau damping is weakly 
irrelevant at the N\'eel quantum critical point, despite the fact that the corresponding self-energy correction dominates over the quadratic gradient terms in the IR limit. 
In the ordered phase, the Landau damping increases under the RG, indicative of damped spin-wave excitations. Although the Kondo coupling is weakly relevant, sufficiently strong Landau damping 
renders the N\'eel quantum critical point quasi-stable for $N_f\ge 4$ and thermodynamically stable for $N_f<4$. In the latter case, we identify a new multi-critical point which describes the transition between 
the N\'eel critical and Kondo run-away regimes. The symmetry breaking at this fixed point results in the opening of a gap in the Dirac fermion spectrum. Approaching the multi-critical 
point from the disordered phase, the fermionic quasiparticle residue vanishes, giving rise to non-Fermi-liquid behavior. 
\end{abstract}

\maketitle

\section{Introduction}

The discovery of topological insulators \cite{Hasan+10,Qi+11} has initiated an explosion of research into Dirac or Weyl semimetals and topological aspects 
of electronic bandstructures \cite{Vafek+14,Yan+17,Armitage+18,Hasan+21}. 
Dirac fermions with relativistic dispersion around point-like Fermi surfaces can arise as low-energy excitations of weakly interacting electron systems. 
The most prominent example is graphene \cite{Neto+09}, which can be described by a tight-binding model of electrons on the half-filled honeycomb lattice. 

Because of their point-like Fermi surfaces nodal semi- metals provide the simplest setting to study fermionic quantum criticality. While Dirac semimetals are 
stable against weak repulsive interactions, a consequence of the vanishing density of states at the Fermi level, sufficiently strong short-range interactions can give rise to a range of 
competing instabilities. For the extended Hubbard model on the half-filled honeycomb lattice rich phase diagrams were 
established \cite{Grushin+13,Garcia+13,Daghofer+14,Capponi+15,Motruk+15,Scherer+15,Volpez+16,Kurita+16,Pena+17,Christou+18}, showing antiferromagnetic, charge ordered, 
Kekule and topological Haldane phases.  Sufficiently strong on-site Hubbard repulsion gives rise to a transition to an antiferromagnetic state with a gap in the electron spectrum 
that is proportional to the staggered magnetization. 

Since the fermionic particle-hole excitations are gapless at such quantum phase transitions, the critical behavior 
falls outside the Landau-Ginzburg-Wilson paradigm of a pure order parameter description \cite{Li+17}. Instead, the nature of the transitions can be studied within a field theory 
that describes the coupling of the bosonic order parameter field, which is introduced through a Hubbard-Stratonovich decoupling of the interaction vertex, to the gapless Dirac 
fermions \cite{Herbut06,Herbut+09,Assaad+13}.   In the high-energy community, such field theories are known as Gross-Neveu-Yukawa (GNY) theories \cite{Gross+74,ZinnJustin91}. 
For the antiferromagnetic transition driven by local Hubbard repulsion the staggered magnetization is described by an $O(3)$ order-parameter field, and the field theory usually referred to as the 
Heisenberg-GNY model. At the fermion-induced GNY fixed point, the fermions acquire an anomalous dimension, resulting 
in the fermion spectral functions with branch cuts rather than quasi-particle poles \cite{Herbut+09}. Such non-Fermi liquid behavior is the hallmark of fermionic quantum criticality. 

Interesting criticality is also expected if Dirac fermions are coupled to local magnetic moments. In the case of graphene, local moments can be introduced by adatoms \cite{Lehtinen+03} 
or defects \cite{Chen+09}, as evidenced by the experimental observation of a Kondo effect \cite{Chen+11}. Moreover, the high Kondo temperatures \cite{Chen+11} reflect the strong 
coupling between the local moments and the conduction electrons in Dirac materials \cite{Hentschel+07}. In subsequent work \cite{Uchoa+11} the unusual Kondo quantum criticality of magnetic 
adatoms in graphene and the very fast power-law decay of the RKKY interaction between them was established. 

In this paper we consider two-dimensional quantum antiferromagnets with Kondo coupling between the local moments and Dirac electrons. While two-dimensional Kondo lattice models 
with Dirac points close to the Fermi level might be rare, there is prospect of engineering such models in heterostructures of graphene and two-dimensional van der Waals magnets \cite{Burch+18}, 
such as the honeycomb Heisenberg antiferromagnets MnPS$_3$ and NiPS$_3$ \cite{Joy+92,Chandrasekharan+94,Wildes+98,Wildes+06,Wildes+15,Lancon+18,Kim+19}.

The aim of this work is to investigate the stability of the N\'eel quantum critical point of a local moment antiferromagnet against the Kondo coupling to Dirac fermions.  We will demonstrate 
that the universal critical behavior of this model is different from that of the Heisenberg GNY theory which describes the antiferromagnetic ordering transition of a purely electronic model 
with local Hubbard repulsion between Dirac electrons.

The outline of this paper is as follows. In Sec.~\ref{sec.model} we introduce the non-linear $\sigma$ model (NL$\sigma$M), describing the long-wavelength behavior of the two-dimensional quantum 
antiferromagnet, with Kondo coupling to $N_f$ Dirac fermion pairs. We discuss the importance of Landau damping of the N\'eel order parameter fluctuations by low-energy electronic particle-hole 
fluctuations. In Sec.~\ref{sec.dampedNLsM} we analyze the Landau-damped  NL$\sigma$M.  Using  momentum-shell RG, we demonstrate that Landau damping is weakly irrelevant at the N\'eel quantum critical 
point but increases in the ordered state, indicating that spin-wave excitations are damped.  The full set of RG equations,  including the Kondo  coupling,  are derived in 
Sec.~\ref{sec.Yukawa} and analyzed in Sec.~\ref{sec.RGflow}. We show that while the Kondo coupling is weakly relevant at the N\'eel quantum critical point, sufficiently strong Landau 
damping renders the critical point quasi-stable for any realistic system size for $N_f\ge 4$ and thermodynamically stable for $N_f<4$. In the latter case, a new multi-critical point captures the transition between 
N\'eel critical and Kondo run-away regimes. We analyze the universal critical behavior associated  with this fixed point.  As demonstrated in Sec.~\ref{sec.eps}, the behavior in $D=3$ space-time dimensions 
is not accessible within an $\epsilon$-expansion above the lower critical dimension, $D=2+\epsilon$.   Finally, in Sec.~\ref{sec.disc} we summarize and discuss our results.

\section{Model}
\label{sec.model}

Our starting model is a non-linear $\sigma$ model (NL$\sigma$M) which describes the N\'eel transition of a two-dimensional quantum antiferromagnet \cite{Chakravarty+88,Chakravarty+89}. This model is coupled to $N_f$ copies of two component 
Dirac electrons via the conventional  Kondo  coupling. On a microscopic level this model could be realized in the low-energy limit of  a quantum antiferromagnet on the honeycomb lattice with  Kondo 
coupling to noninteracting electrons that move on either the same or adjacent honeycomb lattice at half-filling. For this realization we would have $N_f=4$ due to twofold spin and valley degeneracies.  
The effective continuum field theory at zero temperature  is given by the imaginary-time path integral over the action $S = S_f+S_N+S_K$, with contributions
\begin{eqnarray}
\label{eq.action}
S_f & = & \int_{\bq,\omega} \overline{\bm{\psi}}(\bq,\omega)\left(-i\frac{\omega}{v_F} +q_x\tau_x +q_y\tau_y\right)\bm{\psi}(\bq,\omega),\nn\\
S_N & =  & \frac{1}{2g} \int d^2\br \int_0^\infty d\tau\left\{ (\nabla \vec{N})^2+\frac{1}{c^2} (\partial_\tau\vec{N})^2   \right\},\nn\\
S_K & = & \frac{\lambda}{\sqrt{N_f}}  \int d^2\br \int_0^\infty d\tau  \overline{\bm{\psi}} \left( \vec{N}\cdot\vec{\sigma}\otimes \tau_z \right) \bm{\psi},
\end{eqnarray}
where $S_f$ describes two-dimensional Dirac fermions with Fermi velocity $v_F$, written in terms of fermionic Grassmann fields $\bm{\psi}$. The term $S_N$ is the 
conventional NL$\sigma$M in terms of the staggered three-component N\'eel order parameter field $\vec{N}(\br,\tau)$ which satisfies the constraint $\vec{N}^2(\br,\tau)=1$. Here $c$ denotes 
the spin-wave velocity and the coupling constant $g$ is inversely proportional to the spin stiffness. The last contribution  $S_K$ is the Kondo coupling  between the local moments and Dirac electrons. Here 
$\vec{\sigma}$ is the vector of spin Pauli matrices, while the Pauli matrices $\tau_\alpha$ act on sub-lattice space. Note that since $\vec{N}$ describes the staggered magnetization the coupling 
has opposite sign on the two sublattices, resulting in the additional $\tau_z$. The low-energy continuum field theory is subject to a UV momentum cut-off, $|\bq|\le\Lambda$. 

As pointed out in the context of GNY theories \cite{Isobe+16,Uryszek+20},  the Landau damping of the bosonic order parameter fluctuations by electronic particle-hole fluctuations 
gives rise to a self-energy contribution 
\begin{equation}
\Pi(\bq,\omega) = \gamma \sqrt{\bq^2+\omega^2/v_F^2}
\end{equation}
to the inverse boson propagator in two spatial dimensions. This non-analytic self-energy correction 
arrises from the diagram in Fig.~\ref{figure1}(a) from integration of fermion modes near zero momenta and frequency. It is therefore not generated within the momentum shell RG, but needs to be included 
to correctly capture the universal critical behavior of GNY theories  \cite{Isobe+16,Uryszek+20}. The form of the Landau damping does not depend on the number of order parameter components and is not affected
by the fixed-length constraint of the N\'eel order parameter field. Note that although the bare Landau damping parameter $\gamma_0$ is determined by the square of the bare Kondo  coupling, 
$\gamma_0\sim \lambda_0^2$, this relation is not preserved under the RG. We therefore treat $\gamma$ and $\lambda$ as independent coupling constants. 

Under the RG there will be a non-trivial flow of the velocities $c$ and $v_F$. For simplicity, we will focus on the case $v_F=c$, which 
is preserved under the RG. For convenience, we rescale to dimensionless momenta $\bk=\bq/\Lambda$ and frequencies $k_0=\omega/(c\Lambda)$ and absorb the additional prefactors in a 
redefinition of the coupling constants. Since both the order parameter and fermion sectors are relativistic and frequency and momenta enter the zero-temperature field theory 
in the same way, the quantum critical behaviour will be described by a dynamical exponent $z=1$. We will therefore treat frequency and momenta on an equal footing and impose an isotropic cut-off 
in 2+1 dimensions, $\sqrt{\bk^2+k_0^2}\le 1$. Note that the universal critical behavior is independent of the choice of the UV cut-off scheme.

\section{Landau damped NL$\sigma$M}
\label{sec.dampedNLsM}

We start by investigating the effects of Landau damping on the N\'eel transition in the case of vanishing  Kondo coupling, $\lambda=0$.  Our starting point is the 
NL$\sigma$M
\begin{equation}
S_N = \frac{1}{2g}\int_k \Omega^{-1}(k)  \vec{N}(k)\cdot\vec{N}(-k),
\end{equation}
where we have defined $k=(\bk,k_0)$ and $\int_k =\int \frac{d k_0}{2\pi} \int \frac{d^2\bk}{(2\pi)^2}$, subject to the cut-off $|k|\le 1$, for brevity, and include the Landau damping $\gamma$ in the inverse propagator, 
\begin{equation}
\Omega^{-1}(k) =  k^2+\gamma |k| = \bk^2 +k_0^2 + \gamma \sqrt{\bk^2+k_0^2}.
\end{equation}

 We follow the conventional treatment \cite{Chakravarty+88,Chakravarty+89,Nelson+77} and decompose $\vec{N}=(\vec{\pi},\sigma)$ and use the constraint $\sigma(\br,\tau)=\sqrt{1-\vec{\pi}^2(\br,\tau)}$ to 
 eliminate $\sigma$ and derive an effective action in terms of the transverse fields $\vec{\pi}$. In the presence of Landau damping, we need to apply the constraint in momentum space, 
$\sigma(k) = \delta(k) -\frac12 \int_q \vec{\pi}(q)\vec{\pi}(k-q)$. This results in the effective action
\begin{eqnarray}
S_N  & = & \frac{1}{2g} \int_k \Omega^{-1}(k) \vec{\pi}(k)\cdot\vec{\pi}(-k) -\frac{\rho}{2}\int_k \vec{\pi}(k)\cdot\vec{\pi}(-k)\nn\\
& & +\frac{1}{16g}  \int_{k_1,\ldots,k_4} \delta(k_1+k_2+k_3+k_4) \left[  \Omega^{-1}(k_1+k_2)\right.\nn\\
& &  \left. +\Omega^{-1}(k_3+k_4) \right] \left[\vec{\pi}(k_1)\cdot \vec{\pi}(k_2)\right] \left[\vec{\pi}(k_3)\cdot \vec{\pi}(k_4)\right],
\end{eqnarray}
where $\rho$ is the density and the corresponding term arises from exponentiation and expansion of $1/(2\sqrt{1-\vec{\pi}^2(\br,\tau)})$ from the path-integral measure \cite{Nelson+77}. 

\begin{figure}[t]
 \includegraphics[width=0.85\linewidth]{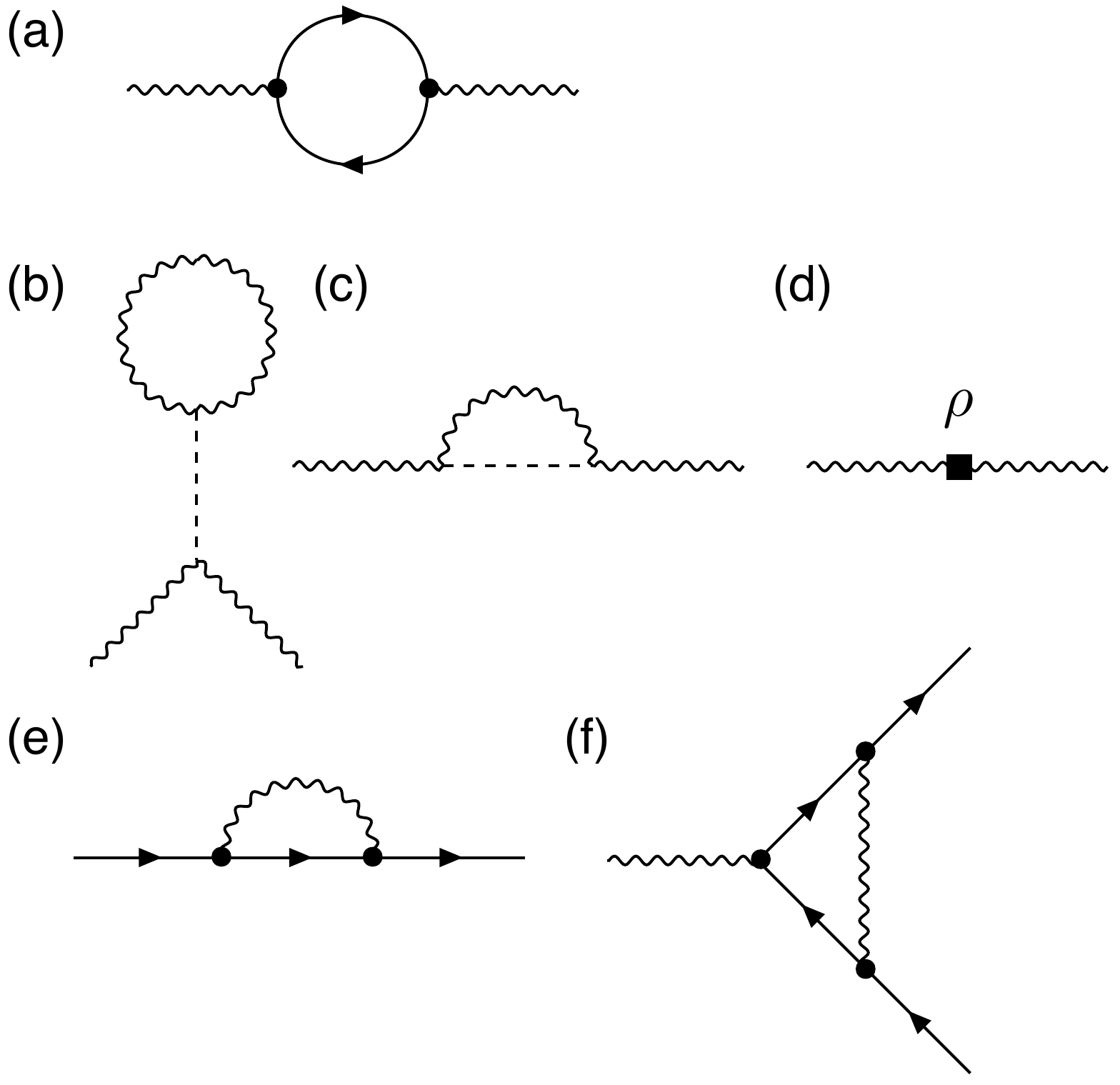}
 \caption{One-loop diagrams relevant to our RG calculation. Solid lines represent fermionic degrees of freedom, wiggly lines the  N\'eel order parameter fields. (a) The fermionic bubble diagram 
 integrated over small momenta and frequencies gives rise to the non-analytic Landau damping of long-wavelength order parameter fluctuations. The momentum-shell contribution of this diagram contributes to 
 the renormalization of the coupling constant $g$ of the NL$\sigma$M. (b)-(d) Diagrams relevant for the RG of the Landau-damped NL$\sigma$M. The diagram (b) is identical to zero, (c) renormalizes the quadratic gradient terms and hence the coupling constant $g$. The unphysical mass term generated by (c) is cancelled by the contribution (d) from the functional integral measure. (e) The fermionic self-energy diagram renormalizes 
 the overall prefactor of the free fermion action $S_f$. The scaling dimension of the fermion fields is determined such that the prefactor remains constant. (f) Diagram contributing to the renormalization of the  Kondo  
 coupling $\lambda$.}
\label{figure1}
\end{figure}

We integrate out modes with momenta and frequencies  from an infinitesimal shell near the cut-off, $e^{-d\ell}\le \sqrt{\bk^2+k_0^2} \le 1$, followed by a rescaling of momenta, $\bk\to\bk e^{d\ell}$, and frequencies,  
$k_0\to k_0 e^{z d\ell}$, with dynamical exponent $z=1$. In addition, we rescale the transverse spin-fluctuation fields as $\vec{\pi}(k)\to \vec{\pi}(k)e^{-\Delta_\pi d\ell}$.

At one-loop order, the contraction of two order parameter fields,
\begin{equation}
\langle \pi_\alpha(k) \pi_\beta(k') \rangle_0 = g \delta_{\alpha\beta}\delta(k+k') \Omega(k),
\end{equation}
 gives rise to the renormalization of the quadratic action by the quartic vertex. The diagram in Fig.~\ref{figure1}(b) vanishes 
because of $\Omega^{-1}(0)=0$. The diagram shown in Fig.~\ref{figure1}(c) gives rise to a term $\sim k^2 \vec{\pi}(k)\cdot\vec{\pi}(-k)$ 
and hence a renormalization of the coupling constant $g$. In addition, it produces a mass term $\sim \vec{\pi}(k)\cdot\vec{\pi}(-k)$ which cancels exactly with the trivial term from the reduction 
of the density $\rho$ by the shell contribution [Fig.~\ref{figure1}(d)]. Evaluating the momentum shell and frequency integrals and combining with the rescaling contributions, we obtain the RG equations
\begin{eqnarray}
\frac{d}{d\ell}\left( \frac{1}{2g} \right) & = & (-5+2\Delta_{\pi} )\frac{1}{2 g} +\frac{1}{(2\pi)^2} \frac{1+\gamma/3}{1+\gamma},\\
\frac{d}{d\ell}\left( \frac{\gamma}{2g} \right) & = &(-4+2\Delta_\pi )\frac{\gamma}{2 g}. 
\end{eqnarray}

The scaling dimension $\Delta_\pi$ of the transverse spin-fluctuation fields needs to be determined such that the constraint of the NL$\sigma$M is satisfied on all scales. This 
is only the case if the coupling constant $g$ in front of the quartic vertex renormalizes in exactly the same way as the $g$ in front of the quadratic action. Instead of evaluating the second-order, one-loop diagrams  
that renormalize the vertex, we employ a trick invented by Nelson and Pelcovits \cite{Nelson+77} to include a staggered magnetic field term $-\frac{h}{2g}\int_{\br,\tau} \sigma(\br,\tau)$ in the action. 
Since the scaling of the magnetic field should not depend on the field direction, and since the staggered magnetic field couples linearly to the N\'eel order parameter field, the scaling dimension of the applied 
field is equal to that of the order-parameter 
field itself, 
\begin{equation}
\frac{d}{d\ell}\left( \frac{h}{2g} \right) = \Delta_\pi  \frac{h}{2g}.
\label{eq.field1}
\end{equation} 

On the other hand, we can use the constraint to expand $\sigma(\br,\tau)$ in terms of the $\vec{\pi}$ fields and explicitly compute the one-loop renormalization of the applied field, 
\begin{equation}
\frac{d}{d\ell}\left( \frac{h}{2g} \right) = (-3+2\Delta_{\pi} )\frac{h}{2 g}+\frac{1}{(2\pi)^2} \frac{h}{1+\gamma}.
\label{eq.field2}
\end{equation}
Equating Eqs.~(\ref{eq.field1}) and (\ref{eq.field2}), we obtain
\begin{equation}
\Delta_{\pi} = 3 - \frac{2}{(2\pi)^2} \frac{g}{1+\gamma},
\label{eq.Dpi}
\end{equation}
which results in the coupled RG equations 
\begin{eqnarray}
\label{eq.RG1}
\frac{d\tilde{g}}{d\ell} & = & -\tilde{g} + \frac{1-\gamma/3}{1+\gamma}\tilde{g}^2,\\
\label{eq.RG2}
\frac{d\gamma}{d\ell} & = & \gamma\left[1-\frac{1+\gamma/3}{1+\gamma}\tilde{g}   \right],
\end{eqnarray}
for $\tilde{g}=\frac{2}{(2\pi)^2} g$ and the Landau damping $\gamma$. 

For $\gamma=0$ we recover the RG equation
of the conventional NL$\sigma$M in 2+1 space-time dimensions. This RG equation exhibits two fixed points: the attractive, N\'eel ordered
fixed point at $\tilde{g}=0$ and the critical fixed point at $\tilde{g}=\tilde{g}_c=1$. For $\tilde{g}(0)<1$ the RG flow is towards $\tilde{g}=0$, corresponding to a freezing 
of transverse spin-fluctuations on larger and larger scales. On the other hand, for $\tilde{g}(0)>1$, $\tilde{g}(\ell)\to\infty$, corresponding to a vanishing of the spin 
stiffness and indicative of the destruction of long-range order by spatial and temporal fluctuations.

\begin{figure}[t]
 \includegraphics[width=0.7\linewidth]{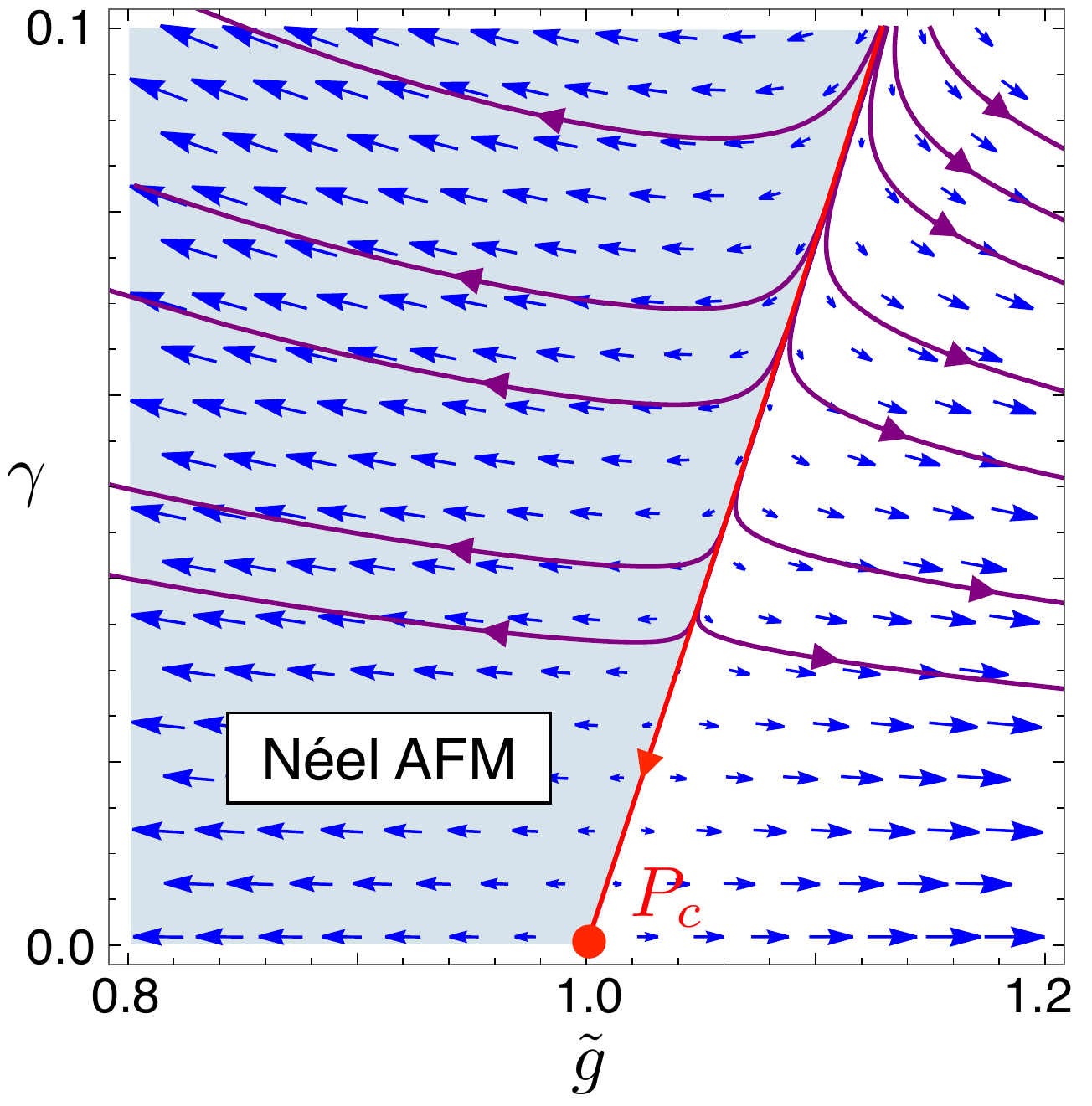}
 \caption{RG flow of the Landau damped NL$\sigma$M as a function of the rescaled inverse spin stiffness $\tilde{g}=g/g_c$ and Landau damping $\gamma$. The red line separates the N\'eel AFM from 
 the paramagnet. Along this separatrix, $\gamma$ renormalizes to zero, demonstrating that the N\'eel quantum critical point $P_c$ is stable against Landau damping. The increase of $\gamma$ in the ordered 
 phase indicates that spin-wave excitations are damped.}
\label{figure2}
\end{figure}

The coupled RG equations (\ref{eq.RG1}) and (\ref{eq.RG2}) do not exhibit any additional fixed points at finite $\gamma$. The RG flow in the $\tilde{g}$-$\gamma$ plane 
is shown in Fig.~\ref{figure2}.  In the antiferromagnetically ordered phase, $\gamma(\ell)$ increases, indicative of damped spin-wave excitations. At the critical fixed point of the N\'eel 
transition the Landau damping $\gamma$ is weakly irrelevant. The separatrix between the N\'eel antiferromagnet and the quantum disordered phase is given by 
$\tilde{g}\approx 1+\frac43 \gamma-\frac49 \gamma^2$. Along the separatrix and for an initial value $\gamma_0=\gamma(0)\ll 1$, the Landau damping vanishes as 
$\gamma(\ell) = \gamma_0/(1+\frac23 \gamma_0 \ell)$.

\section{Including the Kondo coupling to Dirac fermions}
\label{sec.Yukawa}

We now include the  Kondo coupling $S_K$   between the order parameter field to $N_f$ copies of two-component Dirac fermion fields, as given in Eq.~(\ref{eq.action}). The momentum-shell contribution of 
the diagram Fig.~\ref{figure1}(a) will give rise to an additional correction to the NL$\sigma$M,  
\begin{eqnarray}
\delta S_N & = &  -\frac12 \frac{\lambda^2}{N_f} \int_q^<  \vec{N}(q)\cdot\vec{N}(-q) \nn\\
& & \times \int_k^> \textrm{Tr}\left\{\tau_z G_\psi(k)\tau_zG_\psi(k+q)   \right\},
\label{eq.dSN}
\end{eqnarray}
where $\int_q^<$ and  $\int_k^>$ denote frequency-momentum integrals over $|q|\le e^{-d\ell}$ and $e^{-d\ell}\le |k|\le 1$, respectively. The fermionic Green function in each of the $N_f$ copies is given by 
\begin{equation}
G_\psi(k) = \frac{ik_0+k_x \tau_x+k_y \tau_y}{k_0^2+\bk^2}.
\end{equation}

Note that the trace in Eq.~(\ref{eq.dSN}) results in an additional factor of $N_f$. Expanding external momenta/frequencies $q=(\bq,q_0)$ to quadratic order, we obtain
\begin{equation}
\delta S_N =  -\frac13 \lambda^2 \frac{2}{(2\pi)^2}d\ell\int_q^< q^2   \vec{N}(q)\cdot\vec{N}(-q),
\end{equation}
resulting in an additional contribution $d\left(\frac{1}{2g}\right) =  -\frac13 \lambda^2 \frac{2}{(2\pi)^2}d\ell$ to the renormalization of the coupling constant. This changes the RG equations for $\tilde{g}$ and $\gamma$
to
\begin{eqnarray}
\label{eq.RG3}
\frac{d\tilde{g}}{d\ell} & = & -\tilde{g} +\left[ \frac{1-\gamma/3}{1+\gamma}+\frac23 \lambda^2\right]\tilde{g}^2,\\
\label{eq.RG4}
\frac{d\gamma}{d\ell} & = & \gamma\left[1-\frac{1+\gamma/3}{1+\gamma}\tilde{g} +\frac13 \lambda^2\tilde{g}  \right].
\end{eqnarray}

In order to determine the renormalization of the  Kondo coupling constant $\lambda$, we first need to determine the scaling dimension $\Delta_\psi$ of the fermion fields.  The diagram in Fig.~\ref{figure1}(e)
results in a correction 
\begin{eqnarray}
\delta S_f & = & -\frac{2\lambda^2 g}{N_f} \int_k^< \overline{\psi}(k)\left( \int_q^> D(q)\tau_zG_\psi(k+q)\tau_z\right)\psi(k)\nn\\
& = & \frac{2}{3N_f} \frac{1}{1+\gamma}\lambda^2 \tilde{g}\,d\ell  \int_k^< \overline{\psi}(k)G_\psi^{-1}(k)\psi(k),
\end{eqnarray}
where the factor of two arises from the number of components of the transverse spin-fluctuation field $\vec{\pi}$, $N_\pi=2$. \

After rescaling frequency and momenta as before and fermion 
fields as $\psi(k)\to \psi(k) e^{-\Delta_\psi d\ell}$, we demand that that the 
prefactor of $S_f$ remains scale invariant, which results in
\begin{equation}
\Delta_\psi = 2 -\frac{1}{3N_f} \frac{1}{1+\gamma}\lambda^2 \tilde{g}.
\label{eq.Dpsi}
\end{equation}  

The diagram that contributes to the renormalization of the Kondo  vertex is shown in Fig.~\ref{figure1}(f) and equals
\begin{equation}
\delta S_K = \frac{g\lambda^3}{\sqrt{N_f}^3} \sum_i\int_{k_1,k_2}^< \pi_i(k_1-k_2) \overline{\psi}(k_1)\Omega_i \psi(k_2),
\end{equation}
with coupling matrices
\begin{eqnarray}
\Omega_i & = &  \sum_j \int_q^> D(q) (\sigma_j\otimes\tau_z)G_\psi(q)(\sigma_i\otimes\tau_z)\nn\\
& & \times G_\psi(q)(\sigma_j\otimes\tau_z).
\end{eqnarray}

Since $G_\psi$ is independent of spin, we can evaluate the products of spin Pauli matrices and carry out the sum over $j$, 
$\sum_j \sigma_j\sigma_i \sigma_j = (2-N_\pi)\sigma_i$. The momentum-shell integral is trivial and we indeed find that $\Omega_i$
is proportional to the original  Kondo coupling matrix $\sigma_i\otimes\tau_z$, 
\begin{equation}
\Omega_i = \frac{2}{(2\pi)^2}(N_\pi-2)\frac{1}{1+\gamma} d\ell(\sigma_i\otimes\tau_z). 
\end{equation}

However, the result crucially depends upon the number $N_\pi$ of order-parameter components, as discussed in the literature \cite{Sur+19,Uryszek+19}. While the 
results for $N_\pi=1$ and $N_\pi=3$ are equal but of opposite sign, the diagram vanishes in the relevant case of $N_\pi=2$ components,
$\delta S_Y =0$.   

The rescaling of momenta, frequencies and fields gives rise to the RG equation 
\begin{eqnarray}
\label{eq.RG5}
\frac{d\lambda}{d\ell} & = & (-6+\Delta_\pi+2\Delta_\psi)\lambda\nn\\
& = & \lambda \left[1 - \frac{\tilde{g}}{1+\gamma} -\frac{2}{3N_f} \frac{\lambda^2 \tilde{g}}{1+\gamma}\right]
\end{eqnarray}
for the Kondo coupling $\lambda$.

\section{Analysis of RG flow}
\label{sec.RGflow}

We will now discuss the coupled RG equations for the inverse spin stiffness $\tilde{g}$ (\ref{eq.RG3}), the Landau damping $\gamma$ of the N\'eel order parameter (\ref{eq.RG4}) and the 
Kondo  coupling $\lambda$ to the Dirac fermions (\ref{eq.RG5}). In Sec.~\ref{sec.dampedNLsM} we found that in the absence of Kondo  coupling ($\lambda=0$), the Landau damping $\gamma$
is weakly irrelevant at the N\'eel quantum critical point $P_c$. 

\begin{figure}[t]
 \includegraphics[width=0.7\linewidth]{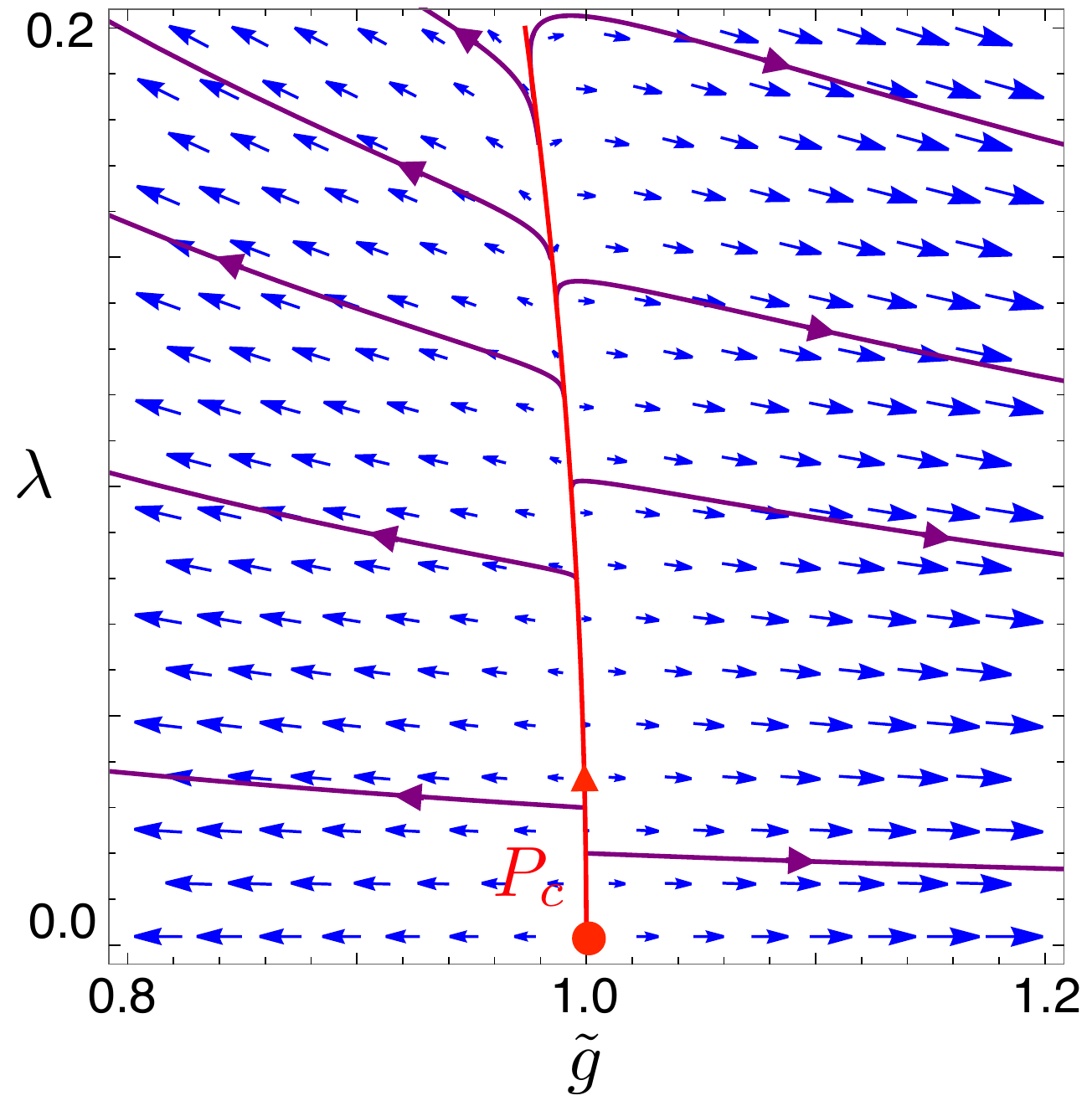}
 \caption{RG flow as a function of inverse spin stiffness $\tilde{g}=g/g_c$ and  Kondo coupling $\lambda$ to Dirac fermions with $N_f=4$ flavors. At the N\'eel quantum critical point $P_c$, the  Kondo  coupling
 is a weakly relevant perturbation, indicated by the increase of $\lambda$ along the separatrix shown in red.}
\label{figure3}
\end{figure}

Let us first investigate the stability of $P_c$ against  Kondo coupling in the absence of Landau damping ($\gamma=0$). In this case the RG equations reduce to $d\tilde{g}/d\ell  =  -\tilde{g} +\left( 1+\frac23 \lambda^2\right)\tilde{g}^2$ and $d\lambda/d\ell =  \lambda \left(1 - \tilde{g}-\frac{2}{3N_f}\lambda^2 \tilde{g}\right)$. In this case we find a separatrix $\tilde{g}=1-\frac23 \lambda^2$, along which the flow of the Kondo  coupling increases 
according to $d\lambda/d\ell = \frac23 (1-1/N_f)\lambda^3$, resulting in $\lambda(\ell)=\lambda_0/\sqrt{1-\frac43 \lambda_0^2 (1-1/N_f)\ell}$. The N\'eel quantum critical point is therefore very weakly unstable against 
the  Kondo  coupling to Dirac fermions. The RG flow in the $\tilde{g}$-$\lambda$ plane is shown in Fig.~\ref{figure3}.

To determine the critical surface in the three-dimensional parameter space of $\tilde{g}$, $\gamma$ and $\lambda$, we insert a polynomial ansatz $\tilde{g}=f(\gamma,\lambda)$ into the RG equations
(\ref{eq.RG3}), (\ref{eq.RG4}) and (\ref{eq.RG5}). To second order we obtain
\begin{equation}
\label{eq.critical}
\tilde{g} = 1+\frac43 \gamma-\frac49 \gamma^2-\frac23 \lambda^2.
\end{equation}

\begin{figure}[t]
 \includegraphics[width=\linewidth]{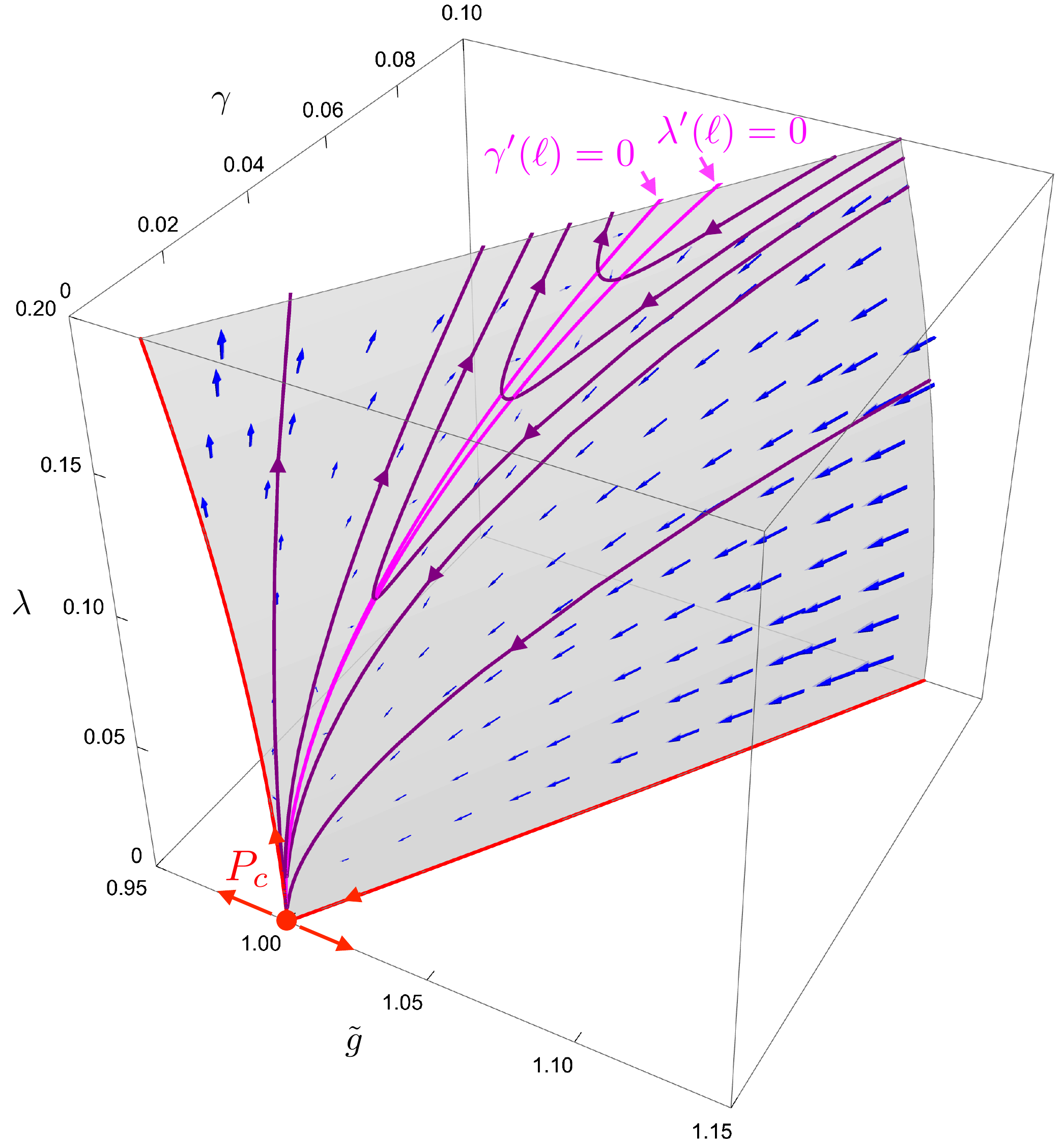}
 \caption{RG flow within the critical surface for $N_f=4$, relevant to Dirac electrons on the honeycomb lattice. The N\'eel quantum critical point $P_c$ is stable against Landau damping 
 $\gamma$ but unstable against  Kondo coupling $\lambda$. For sufficiently strong Landau damping, the RG flow is towards $P_c$ until the trajectories 
 turn to hit the magenta lines, which are given by $\gamma'(\ell)=0$ and $\lambda'(\ell)=0$, respectively, and closely track each other. At this point the RG flow becomes extremely slow and the
  parameters acquire small metastable values.}
\label{figure4}
\end{figure}

The critical surface is shown in Fig.~\ref{figure4}. As expected, the critical surface contains the separatrices in the $\lambda=0$ and $\gamma=0$ planes. For initial values of the coupling constants 
slightly outside the surface, the RG flow is away from the surface: the inverse spin stiffness $\tilde{g}$ renormalizes to zero on one side, indicative of a freezing of spin-wave fluctuations, and to 
infinity on the other side, corresponding to a quantum disordered state. The Landau damping $\gamma$ has a stabilizing effect on the N\'eel order, while the Kondo  coupling $\lambda$ has a destabilizing 
effect.
               
To analyze the competition between $\gamma$ and $\lambda$ within the critical surface we replace $\tilde{g}$ in the corresponding RG equations, using Eq.~(\ref{eq.critical}),
\begin{eqnarray}
\label{eq.RG6}
\frac{d\gamma}{d\ell} & = & -\frac23 \gamma^2+\frac23 \gamma^3+\gamma\lambda^2\\
\label{eq.RG7}
\frac{d\lambda}{d\ell} & = & -\frac13 \gamma\lambda+\frac79 \gamma^2\lambda +\frac23 (1-1/N_f)\lambda^3,
\end{eqnarray}
where we have expanded up to cubic order in the coupling constants. For $N_f\ge 4$, the RG equations only exhibit
a single fixed point at $\lambda=0$ and $\gamma=0$, corresponding to the N\'eel quantum critical point $P_c$. The RG flow in the critical surface and several trajectories 
obtained from numerical integration of the RG equations (\ref{eq.RG6}) and (\ref{eq.RG7}) are shown in Fig.~\ref{figure4} for the case $N_f=4$.

The RG flow is best understood in terms of the lines along which (i) $\gamma'(\ell) = 0$
and (ii) $\lambda'(\ell)=0$, shown in magenta in Fig.~\ref{figure4}, and given by (i) $\lambda^2 = \frac23 (\gamma-\gamma^2)$ and (ii) $\lambda^2 =\frac{1}{2(1-1/N_f)}(\gamma-\frac73 \gamma^2)$, 
respectively. These lines merge at $P_c$ and because of they exhibit the same asymptotic functional form, $\lambda \sim \sqrt{\gamma}$, they closely track each other. 
As a  result, the RG flow becomes very slow in the vicinity of this pair of lines and it is not possible for trajectories to cross them on scales relevant to any realistic system size.

The case $N_f=4$, relevant to Dirac electrons on the honeycomb lattice, is the most extreme since in this case the coefficients of the leading $\sqrt{\gamma}$ terms are identical. 
For weak Landau damping, $\lambda^2 > \frac23 \gamma$, corresponding to points above the magenta lines, the flow is towards the regime of strong Kondo  coupling. This indicates 
that the N\'eel quantum critical $P_c$ point becomes unstable toward Kondo physics, which falls outside the validity of our analysis. 

\begin{figure}[t!]
 \includegraphics[width=0.7\linewidth]{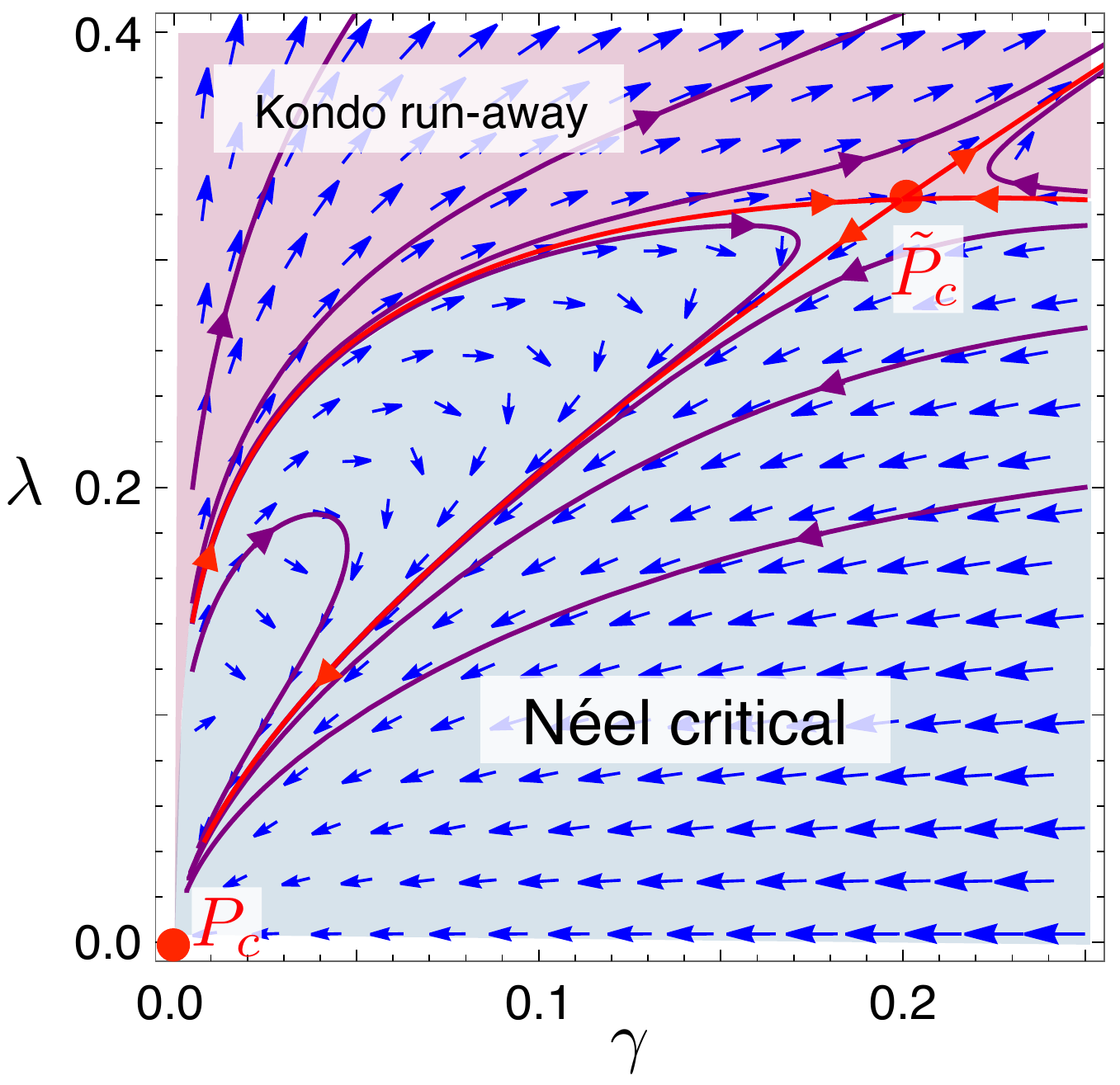}
 \caption{RG flow of the Landau damping $\gamma$ and the Kondo  coupling $\lambda$ within the critical surface for $N_f=2$. The N\'eel quantum critical point $P_c$ is thermodynamically stable in 
 the blue region. In the purple region the flow is towards increasing $\lambda$, indicative of strong coupling Kondo physics. The transition between the two regimes is controlled by a new
 multi-critical point $\tilde{P}_c$.}
\label{figure5}
\end{figure}

On the other hand, if the Landau damping is sufficiently strong, 
$\lambda^2 < \frac23 \gamma$, both $\lambda(\ell)$ and $\gamma(\ell)$ decrease under the RG. The corresponding trajectories approach $P_c$ until they eventually turn to hit the magenta lines. 
Here the RG flow practically comes to a standstill and $\lambda(\ell)$ and $\gamma(\ell)$ reach metastable plateaux values $\gamma_*$ and $\lambda_*^2\approx \frac23 \gamma_*$. 
The non-zero  Kondo  coupling leads to the opening of a small electronic gap $\Delta  \sim\lambda_* | \langle \vec{N}(\mathbf{r},\tau)\rangle |$ in the 
N\'eel ordered phase where the spin-rotational symmetry is broken. On the critical surface the finite values $\gamma_*$ and $\lambda_*$ result in an anomalous contribution to the 
scaling dimension $\Delta_\psi$ of the fermion fields, giving rise to non-Fermi liquid behavior. However, the corresponding critical exponents are non-universal since the 
behavior is not associated with a true fixed point. 

For $N_f<4$ the RG equations (\ref{eq.RG6}) and (\ref{eq.RG7}) exhibit an additional fixed point $\tilde{P}_c$ at 
\begin{equation}
\tilde{\gamma}_c = \frac{4-N_f}{4+3 N_f}, \quad \tilde{\lambda}^2_c = \frac{8}{3}\frac{N_f(4-N_f)}{(4+3N_f)^2},
\label{eq.Pmc}
\end{equation}
which merges with the N\'eel quantum critical point $P_c$ as $N_f\to 4$, showing again that the case $N_f=4$ is marginal. 

In Fig.~\ref{figure5} the RG flow of $\gamma$ and $\lambda$ within the critical surface is shown for the representative case $N_f=2$. In the blue region the RG flow is towards $P_c$, demonstrating 
that the N\'eel fixed point is thermodynamically stable rather than metastable.  In the regime of small  Kondo  coupling $\lambda$, this stability is achieved by finite but very small Landau damping 
$\gamma$. In the purple region the RG flow is towards large values of $\gamma$ and $\lambda$, beyond the validity of our RG equations. The transition between this Kondo run-away regime and the 
N\'eel critical region is described by the multi-critical fixed point $\tilde{P}_c$. 
 
We proceed to analyze the universal critical behavior of the multi-critical point $\tilde{P}_c$ for general $N_f<4$. The correlation length exponent $\tilde{\nu}$ can be obtained from linearizing the RG equation 
for the inverse spin stiffness (\ref{eq.RG3}) around the critical value $\tilde{g}_c\approx1+\frac43 \tilde{\gamma}_c-\frac23\tilde{\lambda}_c^2$. The resulting RG equation is of the general form 
$d(\tilde{g}-\tilde{g_c})/d\ell = \tilde{\nu}^{-1} (\tilde{g}-\tilde{g_c})$. A short calculation gives $\tilde{\nu}=1$ which is identical to the correlation-length exponent $\nu=1$ at the N\'eel quantum critical 
point $P_c$.  

From the scaling dimension $\Delta_\pi$ (\ref{eq.Dpi}) of the transverse spin fluctuations fields $\vec{\pi}$ we obtain the anomalous dimension $\eta_\pi=1$ at the N\'eel critical point $P_c$ and 
\begin{equation}
\tilde{\eta}_\pi = \frac{\tilde{g}_c}{1+\tilde{\gamma}_c}\approx  1+\frac19 \frac{(4-N_f)(12-7 N_f)}{(4+3 N_f)^2}
\end{equation} 
at the new multi-critical point $\tilde{P}_c$. The additional contribution to $\tilde{\eta}_\pi$ results in a slightly different exponent of the algebraic order parameter correlations at criticality, 
$\langle \vec{\pi}(\br)\vec{\pi}(\mathbf{0})\rangle \sim r^{-D+2-\tilde{\eta}_\pi}$, and corrections to other critical exponents, which can be obtained from the conventional scaling and hyper-scaling relations. 

Due to the finite value $\tilde{\lambda}_c$ (\ref{eq.Pmc}) of the Kondo  coupling at $\tilde{P}_c$, the symmetry breaking transition will be accompanied with the opening of a gap \cite{Herbut+09}
\begin{equation}
\Delta\sim (\tilde{g}_c-\tilde{g})^{z\tilde{\nu}}=(\tilde{g}_c-\tilde{g})
\end{equation}
in the Dirac fermion spectrum for $\tilde{g}<\tilde{g}_c$, in the N\'eel ordered phase. Moreover, at $\tilde{P}_c$ the fermions acquire a small anomalous dimension [see Eq.~(\ref{eq.Dpsi})],
\begin{equation}
\tilde{\eta}_\psi = \frac{1}{3N_f}\frac{\tilde{\lambda}_c^2\tilde{g}_c}{1+\tilde{\gamma}_c}\approx \frac89 \frac{4-N_f}{(4+3 N_f)^2},
\end{equation}
which implies that the fermion Green’s function has branch cuts rather than quasiparticle poles. The multicritical point $\tilde{P}_c$ is therefore associated with non-Fermi liquid behavior. 
From a scaling analysis of the fermionic spectral function \cite{Herbut+09} we find that the quasiparticle pole strength vanishes as 
\begin{equation}
Z\sim (\tilde{g}-\tilde{g}_c)^{(z-1+\tilde{\eta}_\psi)\tilde{\nu}} =  (\tilde{g}-\tilde{g}_c)^{\tilde{\eta}_\psi}
\end{equation}
as the critical point is approached from the semi-metallic, non-magnetic phase ($\tilde{g}>\tilde{g}_c$).

\section{Comparison with the $\epsilon$-expansion}
\label{sec.eps}

We now address the question whether the same qualitative behavior can be found within an $\epsilon$-expansion above the lower critical dimension, $D=2+\epsilon$. Although such 
an expansion gives analytic control of the criticality of the NL$\sigma$M \cite{Nelson+77,Polyakov75}, the Pad\'e-Borel  extrapolation to $\epsilon=1$ may be problematic
due to a lack of sign oscillations in the coefficients of the $\epsilon$-expansion of $1/\nu$ \cite{Hikami+78}. 

An additional problem arrises when the NL$\sigma$M is coupled to Dirac fermions since the form of the resulting Landau damping of the N\'eel order parameter field explicitly depends on the 
dimension $D$, $\Pi(\bk)= \gamma |\bk|^{D-2}$ \cite{Uryszek+20}. At one-loop order the RG equations for the inverse spin stiffness $\tilde{g}=g/(2\pi)$ and the Landau damping $\gamma$ in $D=2+\epsilon$ are given by 
\begin{eqnarray}
\frac{d\tilde{g}}{d\ell} & = & -\epsilon\tilde{g}+\frac{1-\gamma \epsilon^2/4}{1+\gamma}\tilde{g}^2,\\
\frac{d\gamma}{d\ell} & = & \gamma\left[2-\epsilon- \frac{1+\gamma \epsilon^2/4}{1+\gamma}\tilde{g} \right],
\end{eqnarray}
where we have determined the scaling dimension of the order parameter field $\Delta_\pi = 2+\epsilon -\tilde{g}/(1+\gamma)$ from the renormalization of an auxiliary magnetic field, as before. 

Without Landau damping, $\gamma=0$, we obtain the N\'eel quantum critical point at $\tilde{g}_c=\epsilon$. At the critical spin stiffness the linearized RG equation for $\gamma$ is equal to 
$d\gamma/d\ell = 2(1-\epsilon)\gamma$, showing that near the lower critical dimension the Landau damping is a relevant perturbation.  

Interestingly, the shell contribution of the diagram in Fig.~\ref{figure1}(a) is equal to zero in $D=2$ due to a vanishing angular integral. As a result, the  Kondo  coupling $\lambda$ does not contribute to the 
renormalization of $\tilde{g}$ and $\gamma$, unlike in $D=3$. 

For similar reasons, the angular integration over the $D=2$ dimensional shell causes the fermionic self-energy diagram, shown in Fig.~\ref{figure1}(e), to vanish. The fermion field does therefore not acquire 
an anomalous dimension and the scaling dimension is trivial, $\Delta_\psi = (3+\epsilon)/2$.  From the scaling dimensions $\Delta_\pi$ and $\Delta_\psi$ we obtain the renormaization of the  Kondo  coupling, 
\begin{equation}
\frac{d\lambda}{d\ell} = \lambda \left[1-\frac{\tilde{g}}{1+\gamma} \right].
\end{equation}

At $\tilde{g}_c=\epsilon$ and $\gamma=0$ the RG equation reduces to $d\lambda/d\ell = (1-\epsilon)\lambda$, demonstrating that the  Kondo coupling is a relevant perturbation at the N\'eel quantum critical point
for $\epsilon<1$. Note that for $\epsilon=1$ both the the Landau damping $\gamma$ and the  Kondo coupling $\lambda$ become marginal, consistent with our calculation in $D=3$.

\section{Discussion}
\label{sec.disc}

We have investigated the stability of the N\'eel quantum critical point of a two-dimensional quantum antiferromagnet with a Kondo coupling to $N_f$ flavors of two-component Dirac fermion fields. For $N_f=4$ this
would describe Dirac electrons on the honeycomb lattice with two-fold spin and valley degeneracies. 

The resulting long-wavelength field theory is given by a NL$\sigma$M  coupled to the Dirac fermion fields. It is crucial to account for the Landau damping of
the N\'eel order parameter field. From simple scaling arguments, the resulting self-energy correction to the order-parameter propagator is expected  to dominate the IR physics. 

At first glance the field theory seems very similar to the Heisenberg GNY theory, which describes the criticality in a purely electronic  model with strong local repulsions between the Dirac electrons.  
There are crucial differences, however. While in the GNY theory the quantum phase transition is tuned by the mass of the order parameter field, the NL$\sigma$M only contains gradient terms and the 
criticality occurs as a function of the inverse spin-stiffness. It is therefore essential to follow the scale dependence of the order parameter propagator with both the quadratic gradient terms and the 
non-analytic self energy correction from Landau damping. In the  Heisenberg GNY theory  on the other hand, the quadratic gradient terms can be discarded.  

Another important difference is that the scaling dimension of the transverse spin-fluctuation field of the NL$\sigma$M is fixed by the requirement 
that the constraint $\vec{N}^2=1$ is satisfied on all length scales. As a result, the boson scaling dimension cannot be used to enforce scale invariance of the  Kondo coupling, in contrast to the 
large-$N_f$ Heisenberg GNY theory.  

We have employed momentum-shell RG to analyze the scale dependence of the inverse spin stiffness $g$, the Landau damping $\gamma$ and the Kondo  coupling $\lambda$. 
Although $\gamma$ and $\lambda$ are initially linked to each other via the fermionic polarization diagram, the two parameters flow independently under the RG. At the N\'eel quantum critical point the scaling 
dimensions of both  $\gamma$ and $\lambda$ vanish and a bifurcation analysis is required. We have investigated the coupled RG flow of the two perturbations within the critical surface $g=f(\gamma,\lambda)$
which contains the unperturbed N\'eel quantum critical point and separates the regions where transverse spin fluctuations freeze or diverge, respectively. 

The flow within the critical surface shows that while the Landau damping $\gamma$ is weakly irrelevant at the N\'eel critical point, the Kondo coupling $\lambda$ is a weakly relevant perturbation. 
Interestingly, the  interplay between the two parameters crucially depends on the number $N_f$ of Dirac fermion flavors. For $N_f\ge 4$, sufficiently strong Landau damping renders the N\'eel  quantum critical point metastable. 
This is evident from an RG flow towards the N\'eel critical point up to scales larger than those relevant to experiments. This behavior is most pronounced for the marginal case $N_f=4$, representing Dirac electrons on the 
honeycomb lattice.

For $N_f<4$ the N\'eel critical point becomes thermodynamically stable over a region where the Landau damping dominates over the Kondo  coupling. We have established a new multicritical point on the 
critical surface which controls the transition between the N\'eel-critical and Kondo-runaway regimes. The finite values of $\gamma$ and $\lambda$ lead to distinct critical exponents and an anomalous dimension of the fermion fields, 
resulting in non-Fermi-liquid behavior. It would be interesting to investigate if our results are robust at higher-loop order and if the change of behavior still occurs at $N_f=4$ or at a different number of Dirac 
fermion flavors. 

Finally, we have investigated the problem in $D=2+\epsilon$ space-time dimensions where the RG calculation of the NL$\sigma$M is controlled. We found that near the lower critical dimension both the Landau damping and the 
 Kondo coupling are relevant perturbations at the N\'eel quantum critical point, resulting in a run-away flow. Unfortunately, the rich behaviour in 2+1 dimensions is not accessible within an $\epsilon$ expansion above the 
lower-critical dimension, and the large-$N_f$ limit does not provide analytic control as in  the Heisenberg GNY theory.  Nevertheless, our results point towards interesting novel critical behavior that 
could potentially be further investigated using quantum Monte Carlo.

\end{document}